\documentclass[twocolumn,pra,showpacs,superscriptaddress,amsmath,amssymb,citeautoscript,aps,10pt]{revtex4-1}
\usepackage{graphicx}
\usepackage{dcolumn}
\usepackage{color}
\usepackage{bm}
\usepackage{braket}
\usepackage[hidelinks]{hyperref}
\usepackage{comment}

\hypersetup{
    colorlinks,
    citecolor=blue,
    filecolor=blue,
    linkcolor=blue,
    urlcolor=blue
}

\usepackage{lipsum}

\allowdisplaybreaks

\begin{document}

%\title{Optimal ultrafast non-adiabatic stitching of a spin chain via universal local control}%: beyound the sudden approximation}

\title{Quantum mechanics and speed limit of ultrafast local control in 
spin chains}

\author{P.~V.~Pyshkin}
\email{pavel.pyshkin@gmail.com}
\affiliation{Institute for Solid State Physics and Optics, Wigner Research Centre for Physics, P.O. Box 49, H-1525 Budapest, Hungary}
\author{E.~Ya.~Sherman}
\affiliation{Department of Physical Chemistry, The University of the Basque Country UPV/EHU, 48080 Bilbao, Spain}
\affiliation{IKERBASQUE, Basque Foundation for Science, 48011 Bilbao, Spain}
\author{Lian-Ao Wu}
\affiliation{Department of Theoretical Physics and History of Science, The University of the Basque Country UPV/EHU, 48080 Bilbao, Spain}
\affiliation{IKERBASQUE, Basque Foundation for Science, 48011 Bilbao, Spain}

\date{\today}

\begin{abstract}
	We study optimization of fidelity for ultrafast transformation of a spin chain via external control of a local exchange coupling. 
	We show that infidelity of such a process can be dramatically decreased by choosing a proper control profile in
	nonadiabatic time domain, predict main features of this profile analytically, corroborate them numerically with 
	a gradient search algorithm, and discuss the corresponding quantum speed limit.
	For ultrafast transformations, the qualitative features of the obtained optimal control are 
	system-independent. Moreover, the main restrictions on its shape do not depend on the transformation time and remain valid up to the
	adiabatic limit. Our results can be applied to control a broad variety of quantum systems.   
\end{abstract}

\maketitle

\section{Introduction}
Recent progress in experimental research on quantum systems described by moderate-size Hilbert spaces, 
such as ensembles of qubits, posed fascinating problems of optimal 
quantum control~\cite{Rabitz-Control-1,control-review-1,Quantum-Control-numerical-book,Caneva-Ref4, Caneva-Ref5} 
of these systems. 
The quantum control aims at achieving desired quantum 
states or certain quantum operations with maximum possible fidelity using limited resources such as time or energy. 
The dynamics of quantum systems under external control can be unitary or non-unitary. 
The unitary dynamics is driven by a time-dependent 
controllable Hamiltonian~$H\left({\mathbf g}(t)\right)$, where~${\mathbf g}(t)$ is a multicomponent control function.
The controllable non-unitary dynamics is achievable by system measurements 
\cite{Li2011,Filippov2017,Wu_entanglement_generation,Pyshkin_compression,measurement-chaos-Kiss-2017},
via a controllable interaction with a non-Markovian 
environment~\cite{Dissipative-control-Verstraete2009,non-markov-Jun-Ting,non-markov-Luo} 
or via control of the unitary part of the evolution of open system~\cite{Dissipative_sys_control}. 

We consider driving a quantum system from a ground state of initial Hamiltonian~$H_{i}$ to achieve at time $T$
ground state of a final Hamiltonian~$H_{f}$ with  $H(t)=H_{i} + g(t)(H_{f}-H_{i}),$ where $g(0)=0$ and $g(T)=1.$ 
Although a high fidelity can be obtained by an adiabatic process~\cite{Born1928} driven by a slowly varying 
$H(t)$ with, e.g., $g(t)=t/T$, this method requires a long evolution while 
optimized~$g(t)$ can permit achieving a demanded quantum state for a relatively short~$T.$

A possible approach to the quantum control, where the transitions occur between the ground states of~$H(t)$,
is based on the shortcut to adiabaticity ~\cite{Shortcuts-Torrontegui-Muga-2013,Muga-review-journal,Demirplak_adiab_drive,Berry-2009,Spin-cutting-PLA-Ren2017}.
However, this technique requires a control of the all parts of a complex quantum system. 
Implementation of a such a shortcut can be a part of quantum computation in arrays of
quantum dots~\cite{Loss1998, Burkard1999}, or in quantum annealing~\cite{Das2008}, such as applied in 
D-Wave computer~\cite{d-wave-book}.  
Here by focusing on high fidelity ultrafast processes, we analytically obtain properties of optimal  {\em local}  control in the ultra-short time domain for a particular many-body system and corroborate our reasoning by a direct 
numerical optimization. We show that several properties of the finite time quantum control (even for the ultra-short time) 
can be explained by requiring a smooth passage to the adiabatic protocols, thus, connecting these two limits.
Although the reported results are obtained for spin chains, 
the proposed heuristic reasoning and numerical approach can be extended to a much broader class of quantum systems.
\begin{figure}
		\includegraphics*[scale=.4]{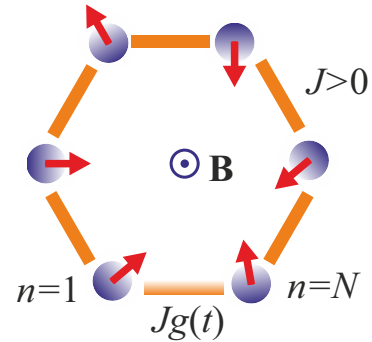}
	\caption{Spin chain with one variable link strength in magnetic field ${\mathbf B}\parallel z-$axis.}
	\label{fig1}
\end{figure}

\section{Ultrafast local control: the problem setting} 
We concentrate on a local control acting only on a small part 
of a complex system, being a natural tool for cutting or stitching 
links between its parts, thus, modifying its size and/or topology. 
We consider an Ising chain with~$N$ spins, as shown in Fig.~\ref{fig1}, described by the Hamiltonian
\begin{equation}\label{ham_{1}}
H(g(t)) = J\sum_{n=1}^{N-1}X_nX_{n+1} + B\sum_{n=1}^{N}Z_n + g(t)JX_{1}X_{N},
\end{equation}
where~$X_{k}, Z_{k}$ are corresponding Pauli matrices of the ~$k$-th spin and $B$ is a magnetic field.
It is useful to rewrite Hamiltonian~(\ref{ham_{1}}) in a short form: $H(t)=H_{0}+g(t)V,$ where $V\equiv JX_{1}X_{N}.$ 
We assume antiferromagnetic interaction and set~$J=1.$ 
The last term in~(\ref{ham_{1}}) connects the first and last spins in the chain. 
By assuming  $g(0)=0$ and $g(T)=1,$ we perform 
a transformation from an open to a ring-shaped chain via ``stitching'' 
a single link between the spins (see Fig.~\ref{fig1}). 
Now we define initial and final Hamiltonians: $H_{i} \equiv H(g(0))$, and $H_{f} \equiv H(g(T))$. 
The corresponding ground states of these Hamiltonians are $\ket{\varphi_{i}}$ and $\ket{\varphi_{f}}$ : 
$H_{i}\ket{\varphi_{i}} = \varepsilon_{i}\ket{\varphi_{i}}$, and $H_{f}\ket{\varphi_{f}} = \varepsilon_{f}\ket{\varphi_{f}}$. 
The state of the system during the evolution is $\ket{\psi(t)} = U(t)\ket{\psi(0)},$ where
\begin{equation}\label{fundamental-U}
U(t) = \mathcal{T}\exp\left(-i\int_{0}^{t}H(g(s))ds\right),
\end{equation}
$\mathcal{T}$ is the time-ordering operator, and we set ~$\hbar\equiv 1$, and the time unit as $1/J$. 
We assume that the initial state~$\ket{\psi(0)} = \ket{\varphi_{i}}$ and 
study the controlled state-transition process with the following target fidelity $f_{T}$ and infidelity $R_{T}$:
\begin{equation}
f_{T} \equiv |\braket{\varphi_{f} | \psi(T)}|, \quad R_{T} \equiv 1 - f_{T}.
\end{equation}
Adiabatic theorem allows us to have an ideal state-transition protocol:
\begin{equation}\label{ideal_adia}
R_{T}\rightarrow0, \quad {\rm for} \quad T\rightarrow\infty, %\quad {\rm and} \quad g(t) = t/T.
\end{equation}
when $g(t)=t/T$. Note that the protocol (\ref{ideal_adia}) is valid only in
the absence of level crossing for an arbitrary~$g(t)\in(0,1)$. 
In this work we assume that the evolution time belongs to one of three domains: ultrashort one ($T\ll1$), 
short time ($T\sim1$), and adiabatic one ($T\gg1$), where the adiabatic theorem is valid. 
We will concentrate mainly on the physics  of the optimal control in the ultrashort domain and analyze
the general features being common for all three domains. Such ultrafast control can be achieved, e.g., 
by electrical manipulation of the bonds connecting quantum dots on the time scale much shorter than global
change in the magnetic field strength \cite{Burkard1999}.

If ground state of~$H_{i}$ or~$H_{f}$ is degenerate we select~$\ket{\varphi_{i}}$ or/and~$\ket{\varphi_{f}}$ 
from some subspace. In such a case we assume that 
$\ket{\varphi_{i}}$ ($\ket{\varphi_{f}}$) is a non-degenerate ground state of 
Hamiltonian~$H(\delta g)$ ($H(1-\delta g)$) for~$\delta g\rightarrow+0$.

Our task is to find the optimal~$g(t)$ to minimize the 
target infidelity functional~$R_{T}[g(t)]$ for a {\em finite} time~$T$.  
Note that our system~(\ref{ham_{1}}) doesn't have complete 
controllability~\cite{controllability-3-Schirmer2002,Controllability-2} because of locality of our control. 
Although the local control can, in general, be complete (see  Ref. \cite{local_controllability-1}), 
the Hamiltonian~(\ref{ham_{1}}) doesn't satisfy assumptions made in Ref. \cite{local_controllability-1} since $iX_{1}X_{N}$
does not generate a Lie algebra in the subspace of 1st and $N$th spins. 
Therefore, the result of our optimization is the minimal nonzero~$R_{T}.$ 

In order to deal with a function instead of a functional we parameterize as follows,
\begin{equation}\label{parametrize}
g(\mathbf{a}, t) = \frac{t}{T} + a_{1}\sin\left(\frac{\pi t}{T}\right) + a_{2}\sin\left(\frac{2\pi t}{T}\right),
\end{equation} 
with $\mathbf{a}\equiv(a_{1}, a_{2})$. 
Now the target infidelity is a function of two parameters~$R_{T} = R_{T}(a_{1}, a_{2})$.
Parametrization~(\ref{parametrize}) can be considered as a simple particular 
case of chopped-random-basis optimization~\cite{CRAB-2-Ref16,CRAB-3-Ref17}.
Our task is to find optimal~$\mathbf{a}=\mathbf{a}_{\rm opt}$, with $R_{T}(\mathbf{a}_{\rm opt})=\min_{\mathbf{a}}\{ R_{T}(\mathbf{a}) \}$. 
Also we can rewrite statement~(\ref{ideal_adia}) as~$R_{T}\rightarrow 0$ for $T\rightarrow\infty$ and $\mathbf{a}\rightarrow \mathbf{0},$ 
and therefore we expect that~$\mathbf{a}=\mathbf{0}$ is a good starting point for gradient numerical search.

\section{Optimal control for ultrafast evolution}

\subsection{Analytical results for singular control parameters}

It is possible to use the first two terms of Dyson series for approximation of~$U(T)$ when $T\rightarrow0$.
One can write 
\begin{eqnarray}
U(T)&\approx& \exp(-iH_{0}T)(\mathbb{I} - iVG(T)) \\
&&\approx(\mathbb{I} - iVG(T))\exp(-iH_{0}T)  \nonumber, \\
G(t)&=&\int_{0}^{t}g(\mathbf{a},s)ds, \nonumber
\end{eqnarray} 
and $\mathbb{I}$ is the identity operator.
By using %$|\braket{\varphi_{f}|H_{0}|\varphi_{i}}| \propto|\braket{\varphi_{f}|\varphi_{i}}|\equiv f_{T}(0)$ and 
$\braket{\varphi_{f}|V|\varphi_{i}} =\braket{\varphi_{f}|H_{f}-H_{i}|\varphi_{i}}=(\varepsilon_{f}-\varepsilon_{i})f_{0}$, where $f_{0}=\braket{\varphi_{f}|\varphi_{i}}$, 
we obtain that this approximation leads to a quadratic $T-$dependence of the target fidelity
\begin{equation}\label{quadratic}
R_{0}-R_{T}\approx |f_{0}|\alpha({\mathbf a})T^2,
\end{equation} 
where $R_{0}\equiv 1-|f_{0}|$, and $\alpha$ is a coefficient, e.g., $\alpha(\mathbf{0})=(\varepsilon_{i}-\varepsilon_{f})^2/8.$  
The quadratic behavior with $dR_{T}/dT|_{T=0}=0$ is an understandable feature of the  
sudden approximation~\cite{book_sakurai} 
where the initial state remains almost intact after fast change in the Hamiltonian. 
However,~(\ref{quadratic})~being valid only for ~$|\mathbf{a}|\ll T^{-1},$ provides an inefficient optimization and, therefore,
one needs to go beyond this condition. 

A more convenient way to go beyond the simple sudden approximation is to apply the following interaction picture:
\begin{equation}
U(T) \approx e^{-iVG(T)}\left(  \mathbb{I} - i \int_{0}^{T} e^{iVG(t)}H_{0}e^{-iVG(t)}dt  \right), \label{Uint}
\end{equation}
with the validity of the integral expression~(\ref{Uint}) being not explicitly related to the magnitude of~$g(t).$  
For $V\equiv X_{1}X_{N}$ we simplify matrix exponents in~(\ref{Uint}) as:
\begin{equation}\label{sin-cos-1}
e^{\pm iG(t)X_{1}X_{N}} = \mathbb{I}\cos G(t) \pm iX_{1}X_{N}\sin G(t).
\end{equation}
Since only two terms, that is $BZ_{1}$ and $BZ_{N},$  in~$H_{0}$ do not commute with~$\exp(\pm i X_{1}X_{N}G(t)),$
by using algebra of Pauli matrices with 
$(X_{1}X_{N})^2=\mathbb{I}$, $X_{k}Z_{k}X_{k} = -Z_{k}$ and $Z_{k}X_{k} = - X_{k}Z_{k} = iY_{k},$ 
we simplify integral in~(\ref{Uint}) as:
\begin{multline}\label{deriv_rotate}
	\int_{0}^{T} e^{iVG(t)}H_{0}e^{-iVG(t)}dt = \int_{0}^{T} \left[ \vphantom{\frac{1}{1}}H_{0}-B\left(Z_{1}+Z_{N}\right) \right. + \\
	\left. \vphantom{\frac{1}{1}} Be^{iVG(t)}(Z_{1}+Z_{N})e^{-iVG(t)} \right]dt = \left(H_{0}-B(Z_{1}+Z_{N})\right) T +\\ 
	  \vphantom{\frac{1}{1}}B(Z_{1}+Z_{N})\beta_{T}(\mathbf{a}) + 
	  \vphantom{\frac{1}{1}}B(Y_{1}X_{N} + X_{1}Y_{N})\gamma_{T}(\mathbf{a}),
\end{multline}
where
\begin{equation}
\hspace{-0.0cm}\beta_{T}(\mathbf{a}) \equiv \int_{0}^{T}\hspace{-0.0cm}\cos(2G(t))dt, 
\quad\hspace{-0.0cm} \gamma_{T}(\mathbf{a}) \equiv \int_{0}^{T}\hspace{-0.0cm}\sin(2G(t))dt. \label{beta-gamma}
\end{equation}
Now we analyze the integral terms in~(\ref{deriv_rotate}). 
If we assume the control is very weak when for any given $t,$ $G(t)\rightarrow0$, we arrive at Eq.~(\ref{quadratic}). 
The opposite case of a strong control should be considered in more detail. 
To get insight into the evolution, we assume for the moment a pulsed control: $g(t)=g_{1}$, for~$0<t<T/2$ and $g(t)=g_{2}$, for~$T/2<t<T$. 
In this case we have~$G(t) = g_{1}t$, $t<T/2$. 
Part of the last integral in~(\ref{deriv_rotate}) can be written as~
\begin{equation}
\int_{0}^{T/2}\sin(2g_{1}t)dt = \frac{T}{2}\cdot(g_{1}T)^{-1}(1-\cos(g_{1}T)), 
\end{equation}
where we picked out linear proportionality on~$T$ as in the other terms in~(\ref{deriv_rotate}). 
Now we see that the contribution of this integral goes to zero in two limits: 1) $g_{1}T\rightarrow0$ and 2) $g_{1}T\rightarrow\infty$. 
Thus, we have reached an important conclusion that for effective control one must have~$g_{1}T = {\rm const}$ for~$T\rightarrow0$. 
The same conclusion can be made for~$g_{2}$ by analysis of the $T/2<t<T$ interval.

As the next step in our reasoning we require that the optimal control enhances the fidelity, that is: 
\begin{equation}\label{instant}
\lim_{T\rightarrow 0}|\braket{\varphi_{f}|U_{\rm opt}(T)|\varphi_{i}}|\geq|\braket{\varphi_{f}|\varphi_{i}}|.
\end{equation} 
In order to satisfy~(\ref{instant}) we can require $\lim_{T\rightarrow0}G(T)=0$ for optimal control (see~Eq.(\ref{Uint})). 
%Later we show that~$f_{T}(T)\rightarrow f_{T}(0)$ when~$T\rightarrow0$ for strong optimal control. 
%This fact together with~(\ref{Uint}) means that~$\lim_{T\rightarrow0}G(T)=0$ for optimal control. 
This means that for two pulses  one must have $g_{2{\rm opt}} = - g_{1 {\rm opt}}$. 
All these conclusions now can be applied for smooth optimal control function~(\ref{parametrize}) in the following way: 
\begin{equation}\label{a_restrict}
\quad \lim_{T\rightarrow 0} a_{1{\rm opt}} = C_{1}, \quad \lim_{T\rightarrow 0} a_{2{\rm opt}} = \frac{C_{2}}{T},
\end{equation}
where constants $C_{1,2}$ to be obtained by numerical calculations based on Eq.~(\ref{deriv_rotate}).

Using result~(\ref{deriv_rotate}) we write the target infidelity as:
\begin{eqnarray}
&&R_{0}-R_{T} \approx |f_{0}|B\left[\left(\vphantom{1^1}  \beta_{T}(\mathbf{a}) - T\right)F_{1} + \gamma_{T}(\mathbf{a})F_{2} \right], \label{f_linear}\\
&&F_{1} = {\rm Im}\frac{f_{Z}f^{*}_{0}}{|f_{0}|^2}, \quad F_{2} = {\rm Im}\frac{f_{XY}f^{*}_{0}}{|f_{0}|^2},
\end{eqnarray}
where
\begin{equation}
f_{Z} = \braket{\varphi_{f}|Z_{1}+Z_{N}|\varphi_{i}}, f_{XY}=\braket{\varphi_{f}|X_{1}Y_{N}+Y_{1}X_{N}|\varphi_{i}}.
\end{equation}
Since without loss of generality, we can assume that the states $\ket{\varphi_i}$ and  $\ket{\varphi_f}$ are real, we
obtain $F_{1}=0$, ${\rm Re}F_{2}=0$ and thus our result does not depend on~$\beta_{T}(\mathbf{a}).$

Expression (\ref{f_linear}) was derived using
\begin{multline}
\braket{\varphi_{f}|U(T)|\varphi_{i}} = 
\bra{\varphi_{f}} \left[\vphantom{\frac{1}{1}} \mathbb{I} - i B(Z_{1}+Z_{N}) (\beta_T(\mathbf{a})-T)-\right.\\
\left. \vphantom{\frac{1}{1}} iB(Y_{1}X_{N}+X_{1}Y_{N})\gamma_T(\mathbf{a})\right] e^{-iH_{0}T}\ket{\varphi_{i}} + \mathcal{O}(T^2). 
\end{multline} 
Here we also assume~$G(T)=0$ for all values of $T$. This assumption means that we chose~$a_{1} = -\pi/4$ as optimal value 
for any small nonzero~$T$~\cite{2019-1}. 
%In other words we change the real function $G(T)$ which has a property $G(T)\rightarrow0$ for $T\rightarrow0$ to a constant~$G(T)=0$.   
It is important that optimized function~$\gamma_{T}(\mathbf{a}_{\rm opt}) = K_{\gamma} T$ is linear for 
small~$T$, where $\gamma_{T}(\mathbf{a}_{\rm opt}) = \max_{\mathbf{a}}\{ \gamma_{T}(\mathbf{a})\}$, $K_{\gamma}$~is a system-dependent coefficient, and~$\mathbf{a_{{\rm opt}}}$ satisfies~(\ref{a_restrict}).
Thus, the linear approximation to optimal infidelity is 
\begin{equation}\label{f_linear_simple}
R_{0}-R_{T} = |f_{0}|B K_{\gamma} F_{2} T.
\end{equation}
Linearly decreasing behavior of infidelity~(\ref{f_linear_simple}) under optimal control gives a big advantage in comparison with quadratic 
(\ref{quadratic}) for short time~$T$.  The spatial symmetry of Hamiltonian (\ref{ham_{1}}) assures that $F_{2}$ is an odd function of~$B,$ 
corresponding to the fidelity independent of the direction of the magnetic field.

\subsection{Numerical examples} 

\begin{figure}
		\includegraphics*{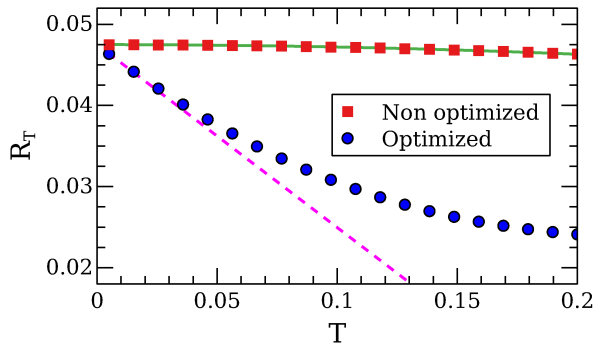}
	\caption{Optimized and non-optimized infidelity~$R_{T}$ for ultrashort timescale as a function of the process time~$T$.
	The linear approximation~(\ref{f_linear_simple}) for optimized infidelity is given by the dashed line. }
	\label{fig2}
\end{figure}

\begin{figure}
	
	\includegraphics*{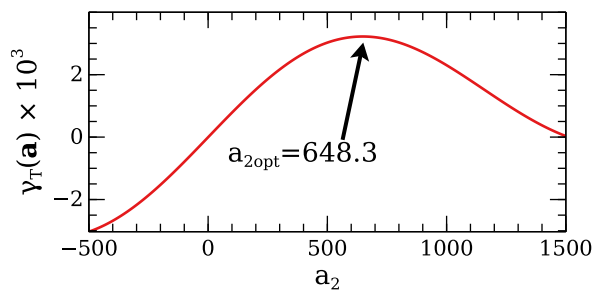}
	
	\caption{Example of the dependence of~$\gamma_{T}(\mathbf{a})$ function for fixed~$T=0.005$ and $a_{1}=-\pi/4$. 
	The optimal parameters from direct numerical BFGS optimization are: $a_{1{\rm opt}}=-0.9$ and $a_{2{\rm opt}}=646.8$.}
	\label{fig3}
\end{figure}

To illustrate the above arguments, we study a chain with~$N=6$ spins and~$B=0.9$, and
relate the results of direct numerical simulations to expressions~(\ref{quadratic}),~(\ref{f_linear}), and (\ref{f_linear_simple}). 
Although the Ising chain in a transverse field is exactly solvable~\cite{Ising-exact}, we 
obtain the states $\ket{\varphi_{i}}$ and $\ket{\varphi_{f}}$ by direct numerical diagonalization of the corresponding 
Hamiltonians. Next, we use gradient Broyden-Fletcher-Goldfarb-Shanno (BFGS) algorithm~\cite{BFGS} in direct numerical search of~$\mathbf{a}_{\rm opt}(T)$. 
Exact numerical diagonalization was used in order to calculate $\mathbf{a}-$dependent propagators~(\ref{fundamental-U}). 
In Fig.~\ref{fig2} we show non-optimized ($g(t)=t/T$) and optimized infidelities obtained by direct numerical simulations. 
%The green solid line corresponded to non-optimized process was achieved from analytical approximation~(\ref{quadratic}). 
To quantify linear approximation for the optimized fidelity, we first numerically 
obtain~$f_{0}=0.9525$, $f_{Z}=-1.4090$, $f_{XY}=0.389i$, resulting in ~$F_{1}=0,$ as expected, and $F_{2} = 0.408$. 
It is easy to numerically find a maximum of a function~$\gamma_T(\mathbf{a})$ for fixed~$T$ and~$a_{1}=-\pi/4$; 
the example of dependence~$\gamma_T(a_{2})$ for~$T=0.005$ is depicted in Fig.~\ref{fig3}.
From the last line of Table~\ref{t1} we find~$K_{\gamma}=0.644$ and obtain $B K_{\gamma} F_{2}=0.237$.
Additional numerical checks show that adding next harmonic in~(\ref{parametrize}) 
doesn't considerably change the optimal fidelity value. Note, that expression~(\ref{f_linear_simple}) being linear in~$T,$ 
is not linear in~$B$ since $F_{2}$ is $B-$dependent.     

\begin{table}[]
	\centering
	\caption{Numerical evidence of linearity of~$\gamma_{T}(\mathbf{a}_{\rm opt})$.}
	%\begin{tabu} to 0.8\textwidth
	\begin{tabular} {l l l l l}
		%\multicolumn{4}{ c } \\ \hline
		\hline 
		\hline
		$T$& 0.005 & 0.02 & 0.05 & 0.1  \\
		$a_{2{\rm opt}}$ &  648.3 & 162.4 & 65.2 & 32.84\\
		$\gamma_{T}(\mathbf{a}_{\rm opt})$  & 0.00322 & 0.0129 & 0.0322 & 0.0642 \\
		$\gamma_{T}(\mathbf{a}_{\rm opt})/ T$ & 0.644 & 0.644 & 0.643 & 0.642 \\ 
		\hline 
		\hline
	\end{tabular}
	\label{t1}
\end{table}

\begin{figure}
	
		\includegraphics*{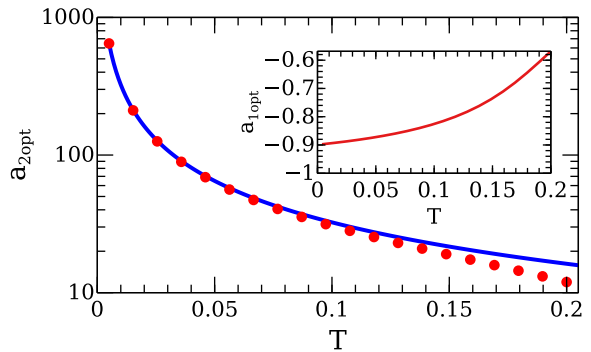}

	\caption{Values of parameters for optimized control for ultrashort timescale as a function of the process time~$T$. Red circles in the main panel and the solid line 
	in the inset are result of exact BFGS optimization, blue solid line in the main panel is analytical approximation~$a_{2}(T)=3.24/T$.  }
	\label{fig4}
\end{figure}

In Fig.~\ref{fig4} (main panel) we show numerically optimized values of~$a_{2{\rm opt}}$, and 
approximation~$a_{2}(T)=3.24/T,$ which follows from expression~(\ref{a_restrict}). 
The coefficient~$3.24$ can be obtained by taking product~$Ta_{2{\rm opt}}$ from the first column of Table~\ref{t1}. 
We see a good agreement between analytical and exact numerical optimization up to~$T\approx0.1$. 
In the inset of~Fig.~\ref{fig4} we show numerically optimized values of~$a_{1{\rm opt}}(T)$ and 
see that~$a_{1}$ remains finite in the limit~$T\rightarrow0$ in agreement with~(\ref{a_restrict}), and 
thus our simplification $G(T)=0$ for~$T\rightarrow0$ in~(\ref{f_linear}) was rational. 
Note, in Fig.~\ref{fig4} we do not have $\lim_{T\rightarrow 0}a_{1{\rm opt}}=-\pi/4$ as we used in our analysis.
However, the approximation we made is valid because $G(T)=T(1/2 + 2a_{1{\rm opt}}/\pi) \ll 1$ for 
any {\em finite}~$|a_{1{\rm opt}}|$ and~$T\rightarrow0$.

Important input of~(\ref{f_linear_simple}) is that the optimal control parameters~$\mathbf{a}_{\rm opt}$ can be evaluated from analysis of 
function~$\gamma_{T}(\mathbf{a})$ (\ref{beta-gamma}), which, in turn, does not contain information about~$N$ and~$B$. 
Therefore we expect that the optimal control~$g(t)$ for different quantum Ising 
chains is only very weakly 
system-dependent in the ultrashort time limit. This {\em universality} is confirmed by Fig.~\ref{fig5} (a) 
where we present the BFGS optimization results for chains with different parameters.  

\subsection{Relation to quantum speed limit}

\begin{figure}
	
	\includegraphics*{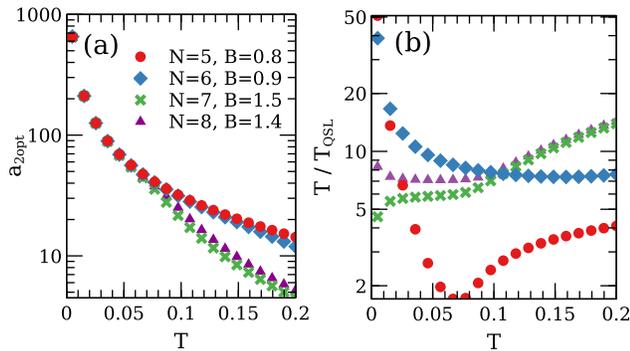}
	
	\caption{(a): The optimal $T-$dependent control parameters for various values of $N$ and $B$. This Figure shows that 
		 in the limit~$T\rightarrow0$ they are almost the same for different spin chains and fields $B$. 
		 (b): The ratios $T/T_{\rm QSL}$ as functions of~$T$ for optimized evolution. }
	\label{fig5}
\end{figure}

The optimal control (\ref{a_restrict}) produces a strong perturbation,
where the characteristic energy given by the time-energy uncertainty \cite{Mandelstam1991} is proportional to~$T^{-1}.$ 
Therefore, it is instructive to consider the quantum speed limit (QSL) time~$T_{\rm QSL},$ that is 
the minimal possible time required to transform the initial $|\varphi_{i}\rangle$ into the final $U_{\rm opt}(T)|\varphi_{i}\rangle$ state
\cite{MARGOLUS1998188}. 
The $T_{\rm QSL}$ time is computed as \cite{QSL_Deffner_2013}:
\begin{equation}
\frac{T_{\rm QSL}}{T} = 
\frac{\displaystyle\arccos\left|\vphantom{1^1}\braket{\varphi_i|U_{\rm opt}(T)|\varphi_i}\right|}
{\displaystyle\int_0^T \left| \vphantom{\frac{1}{1}}\braket{\varphi_i|H(g(\mathbf{a}_{\rm opt},t)|\varphi_i} \right|dt}, 
\end{equation}
and the ratio $T/T_{\rm QSL}$ can be considered as the efficiency of the quantum control 
(see, e.g.,~\cite{QSL-in-chain-Murphy-Ref15,QSL-quant-contr}). 
In~Fig. \ref{fig5}(b), where  we show $T-$dependence of $T/T_{\rm QSL}$, one can see the ratio~$T/T_{\rm QSL}\gtrsim 5$, 
and this relatively large value can be related to the locality of the control.

\begin{comment}
\begin{figure}

		\includegraphics*{universal_control.eps}

	\caption{The optimal $T-$dependent control parameters for various values of $N$ and $B$. This Figure shows that they 
	are almost the same for different spin chains and fields $B$.}
	\label{fig5}
\end{figure}
\end{comment}

\section{From ultrafast to adiabatic evolution} 

\begin{figure}
		\includegraphics*{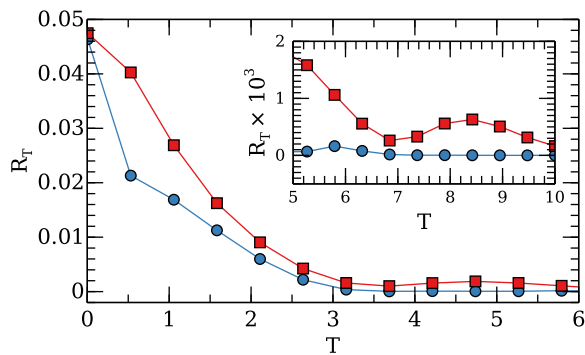}
	\caption{Optimized (circles) and non-optimized (squares) infidelities~$R_{T}$ for long timescale as function of the process time~$T$.}
	\label{fig6}
\end{figure}

Here we briefly discuss features 
of optimal control when the evolution time~$T$ runs from ultrashort values to the adiabatic domain.
In~Fig.~\ref{fig6} we show non-optimized and optimized infidelity for an extended interval of $T.$
As can be seen, the non-optimized infidelity goes to zero as a consequence of adiabaticity~(\ref{ideal_adia}). 
Also, the maximal difference between optimized and non-optimized infidelity appears at short time~$T\lesssim1$.

As we have shown in Refs.~\cite{Pyshkin2018-NJP,Acta2019} the optimal shape of control function~$g(t)$ 
is restricted by two requirements: (1) continuous transition from non-adiabatic to adiabatic 
time domain, and (2) nonzero time derivative~$\dot{g}(t)$ at $t=0$ and $t=T.$  
These two assumptions lead to the following conditions:
\begin{equation}\label{inequal_conditions}
\dot{g}(t\rightarrow 0) > 0, \quad \dot{g}(t\rightarrow T) > 0,
\end{equation}
written in parametrization~(\ref{parametrize}) as:
\begin{equation}\label{inequal_a}
-\left({1}/{\pi} + 2a_{2}\right) < a_{1} <  {1}/{\pi} + 2a_{2}. 
\end{equation}
Remarkably,  the conditions (\ref{inequal_a}) are $T-$independent. It turns out that~$\mathbf{a}_{\rm opt}$ from the numerical calculations 
satisfies~(\ref{inequal_a}) for ultrashort, short, and adiabatic processes, regardless of the time domain
(this statement is as well $T-$independent when we take more than two harmonics in~(\ref{parametrize})). %Lianao: check if the rewritting is what you meant.
Note that first two harmonics in~(\ref{parametrize}) provide the simplest efficient parametrization, 
which satisfies~(\ref{inequal_conditions}), for a strong energy pumping, with~$|g(t)|\gg1$ at a finite time interval.

\begin{figure}
	\includegraphics*{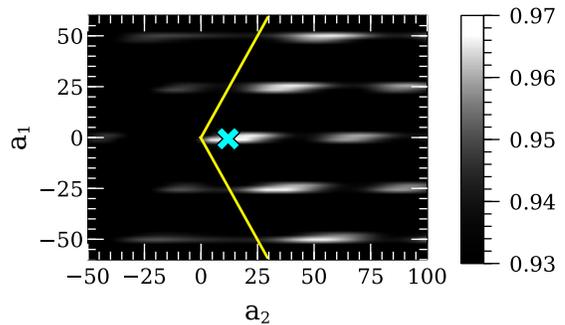}
	\caption{Landscape of output fidelity as a function of~$a_{1}$ and~$a_{2}$ for~$T=0.2$. 
	 Yellow lines correspond to conditions~(\ref{inequal_a}) and cyan cross is the numerically obtained optimal value.}
	\label{fig7}
\end{figure}

In Fig.~\ref{fig7} we show the landscape of output fidelity as a function of~$\mathbf{a}$ for~$T=0.2$. 
One can see that the high-fidelity ``islands'' form horizontal equidistant lines. 
%corresponded to~$a_{1}=\pi^2l/2T\approx 24.7 l$ ($l=0, \pm1, \pm2, \dots$). 
Appearance of these lines is related to possibility of the satisfaction of~(\ref{instant}) by letting~
$U(T\rightarrow0)=\mathbb{I}$ with $a_{1}=\pi^2l/2T$ $(\approx 24.7l,\,l=\pm1, \pm2, \dots)$ (see Eqs.(\ref{Uint}) and~(\ref{sin-cos-1})).
In comparison with the fidelity of different local maximums we see that initial point~$\mathbf{a}=\mathbf{0}$ is the 
valid choice for numerical BFGS search in order to avoid traps~
\cite{easy-control-Rabitz1998,easy-to-find-control,are-there-traps-in-landscapes,role-of-constraints-in-control,easy-control-criticizm-Zhdanov2018},
and our assumption~$\lim_{T\rightarrow 0}G(T)=0$ is corroborated. 
Moreover, the universal initial point~$(0,0)$ (which satisfies~(\ref{inequal_a})) for numerical 
search connects together adiabatic and non-adiabatic time domains.

\section{Conclusion} We demonstrated that properly-designed incomplete local control can greatly decrease infidelity 
of unitary evolution in the non-adiabatic time domain, even for ultrafast transition processes. %Pavel: "incomplete" is more informative than inperfect for quantum-control community
We presented an approximate analytical solution for finding the optimal control parameters in the ultrashort~$T$ 
domain and showed that optimization can lead to a linear in
~$T$ decrease in the infidelity.
Rather than achieving zero infidelity, this linearity is the main benefit of 
using the unrestricted energy resource in the case of incomplete local control.

The main features of the optimal control found by heuristic reasoning and analytical derivations
have been confirmed by direct numerical simulations. Our results show that optimal control parameters for short $T$,  
being system-independent,  are somehow universal.   
Surprisingly, in our approach one needs to only analyze one of extrema of a single-variable analytical function to find the optimal control 
parameters instead of the conventional numerical algorithm for computing propagators. 
We hope that our findings and approaches will be useful for further improvements of efficiency in realistic quantum control
in broad variety of systems.

%%%%%%%%%%%%%%%%%%%%%%%%%%%%%%%%%%%%%

\section*{Acknowledgements} We gratefully acknowledge National Research, Development and Innovation Office of Hungary
(Project Nos. K124351 and 2017-1.2.1-NKP-2017-00001), the Basque Country Government (Grant No. IT472-10), 
the Spanish Ministry of Economy, Industry, and Competitiveness (MINECO) 
and the European Regional Development Fund FEDER Grant No. FIS2015-67161-P (MINECO/FEDER, UE).

%%%%%%%%%%%%%%%%%%%%%%%%%%%%%%%%%%%%%

%\bibliography{biblioteka.bib}

\begin{thebibliography}{45}%
	\makeatletter
	\providecommand \@ifxundefined [1]{%
		\@ifx{#1\undefined}
	}%
	\providecommand \@ifnum [1]{%
		\ifnum #1\expandafter \@firstoftwo
		\else \expandafter \@secondoftwo
		\fi
	}%
	\providecommand \@ifx [1]{%
		\ifx #1\expandafter \@firstoftwo
		\else \expandafter \@secondoftwo
		\fi
	}%
	\providecommand \natexlab [1]{#1}%
	\providecommand \enquote  [1]{``#1''}%
	\providecommand \bibnamefont  [1]{#1}%
	\providecommand \bibfnamefont [1]{#1}%
	\providecommand \citenamefont [1]{#1}%
	\providecommand \href@noop [0]{\@secondoftwo}%
	\providecommand \href [0]{\begingroup \@sanitize@url \@href}%
	\providecommand \@href[1]{\@@startlink{#1}\@@href}%
	\providecommand \@@href[1]{\endgroup#1\@@endlink}%
	\providecommand \@sanitize@url [0]{\catcode `\\12\catcode `\$12\catcode
		`\&12\catcode `\#12\catcode `\^12\catcode `\_12\catcode `\%12\relax}%
	\providecommand \@@startlink[1]{}%
	\providecommand \@@endlink[0]{}%
	\providecommand \url  [0]{\begingroup\@sanitize@url \@url }%
	\providecommand \@url [1]{\endgroup\@href {#1}{\urlprefix }}%
	\providecommand \urlprefix  [0]{URL }%
	\providecommand \Eprint [0]{\href }%
	\providecommand \doibase [0]{http://dx.doi.org/}%
	\providecommand \selectlanguage [0]{\@gobble}%
	\providecommand \bibinfo  [0]{\@secondoftwo}%
	\providecommand \bibfield  [0]{\@secondoftwo}%
	\providecommand \translation [1]{[#1]}%
	\providecommand \BibitemOpen [0]{}%
	\providecommand \bibitemStop [0]{}%
	\providecommand \bibitemNoStop [0]{.\EOS\space}%
	\providecommand \EOS [0]{\spacefactor3000\relax}%
	\providecommand \BibitemShut  [1]{\csname bibitem#1\endcsname}%
	\let\auto@bib@innerbib\@empty
	%</preamble>
	\bibitem [{\citenamefont {Peirce}\ \emph {et~al.}(1988)\citenamefont {Peirce},
		\citenamefont {Dahleh},\ and\ \citenamefont {Rabitz}}]{Rabitz-Control-1}%
	\BibitemOpen
	\bibfield  {author} {\bibinfo {author} {\bibfnamefont {A.~P.}\ \bibnamefont
			{Peirce}}, \bibinfo {author} {\bibfnamefont {M.~A.}\ \bibnamefont {Dahleh}},
		\ and\ \bibinfo {author} {\bibfnamefont {H.}~\bibnamefont {Rabitz}},\
	}\href@noop {} {\bibfield  {journal} {\bibinfo  {journal} {Phys. Rev. A}\
		}\textbf {\bibinfo {volume} {37}},\ \bibinfo {pages} {4950} (\bibinfo {year}
		{1988})}\BibitemShut {NoStop}%
	\bibitem [{\citenamefont {Brif}\ \emph {et~al.}(2010)\citenamefont {Brif},
		\citenamefont {Chakrabarti},\ and\ \citenamefont
		{Rabitz}}]{control-review-1}%
	\BibitemOpen
	\bibfield  {author} {\bibinfo {author} {\bibfnamefont {C.}~\bibnamefont
			{Brif}}, \bibinfo {author} {\bibfnamefont {R.}~\bibnamefont {Chakrabarti}}, \
		and\ \bibinfo {author} {\bibfnamefont {H.}~\bibnamefont {Rabitz}},\
	}\href@noop {} {\bibfield  {journal} {\bibinfo  {journal} {New Journal of
				Physics}\ }\textbf {\bibinfo {volume} {12}},\ \bibinfo {pages} {075008}
		(\bibinfo {year} {2010})}\BibitemShut {NoStop}%
	\bibitem [{\citenamefont {Borzi}\ \emph {et~al.}(2017)\citenamefont {Borzi},
		\citenamefont {Ciaramella},\ and\ \citenamefont
		{Sprengel}}]{Quantum-Control-numerical-book}%
	\BibitemOpen
	\bibfield  {author} {\bibinfo {author} {\bibfnamefont {A.}~\bibnamefont
			{Borzi}}, \bibinfo {author} {\bibfnamefont {G.}~\bibnamefont {Ciaramella}}, \
		and\ \bibinfo {author} {\bibfnamefont {M.}~\bibnamefont {Sprengel}},\
	}\href@noop {} {\emph {\bibinfo {title} {Formulation and Numerical Solution
				of Quantum Control Problems (Computational Science \& Engineering)}}}\
	(\bibinfo  {publisher} {SIAM-Society for Industrial \& Applied Mathematics},\
	\bibinfo {year} {2017})\BibitemShut {NoStop}%
	\bibitem [{\citenamefont {Caneva}\ \emph
		{et~al.}(2011{\natexlab{a}})\citenamefont {Caneva}, \citenamefont {Calarco},
		\citenamefont {Fazio}, \citenamefont {Santoro},\ and\ \citenamefont
		{Montangero}}]{Caneva-Ref4}%
	\BibitemOpen
	\bibfield  {author} {\bibinfo {author} {\bibfnamefont {T.}~\bibnamefont
			{Caneva}}, \bibinfo {author} {\bibfnamefont {T.}~\bibnamefont {Calarco}},
		\bibinfo {author} {\bibfnamefont {R.}~\bibnamefont {Fazio}}, \bibinfo
		{author} {\bibfnamefont {G.~E.}\ \bibnamefont {Santoro}}, \ and\ \bibinfo
		{author} {\bibfnamefont {S.}~\bibnamefont {Montangero}},\ }\href@noop {}
	{\bibfield  {journal} {\bibinfo  {journal} {Phys. Rev. A}\ }\textbf {\bibinfo
			{volume} {84}},\ \bibinfo {pages} {012312} (\bibinfo {year}
		{2011}{\natexlab{a}})}\BibitemShut {NoStop}%
	\bibitem [{\citenamefont {Caneva}\ \emph {et~al.}(2014)\citenamefont {Caneva},
		\citenamefont {Silva}, \citenamefont {Fazio}, \citenamefont {Lloyd},
		\citenamefont {Calarco},\ and\ \citenamefont {Montangero}}]{Caneva-Ref5}%
	\BibitemOpen
	\bibfield  {author} {\bibinfo {author} {\bibfnamefont {T.}~\bibnamefont
			{Caneva}}, \bibinfo {author} {\bibfnamefont {A.}~\bibnamefont {Silva}},
		\bibinfo {author} {\bibfnamefont {R.}~\bibnamefont {Fazio}}, \bibinfo
		{author} {\bibfnamefont {S.}~\bibnamefont {Lloyd}}, \bibinfo {author}
		{\bibfnamefont {T.}~\bibnamefont {Calarco}}, \ and\ \bibinfo {author}
		{\bibfnamefont {S.}~\bibnamefont {Montangero}},\ }\href@noop {} {\bibfield
		{journal} {\bibinfo  {journal} {Phys. Rev. A}\ }\textbf {\bibinfo {volume}
			{89}},\ \bibinfo {pages} {042322} (\bibinfo {year} {2014})}\BibitemShut
	{NoStop}%
	\bibitem [{\citenamefont {Li}\ \emph {et~al.}(2011)\citenamefont {Li},
		\citenamefont {Wu}, \citenamefont {Wang},\ and\ \citenamefont
		{Yang}}]{Li2011}%
	\BibitemOpen
	\bibfield  {author} {\bibinfo {author} {\bibfnamefont {Y.}~\bibnamefont
			{Li}}, \bibinfo {author} {\bibfnamefont {L.-A.}\ \bibnamefont {Wu}}, \bibinfo
		{author} {\bibfnamefont {Y.-D.}\ \bibnamefont {Wang}}, \ and\ \bibinfo
		{author} {\bibfnamefont {L.-P.}\ \bibnamefont {Yang}},\ }\href@noop {}
	{\bibfield  {journal} {\bibinfo  {journal} {Phys. Rev. B}\ }\textbf {\bibinfo
			{volume} {84}},\ \bibinfo {pages} {094502} (\bibinfo {year}
		{2011})}\BibitemShut {NoStop}%
	\bibitem [{\citenamefont {Luchnikov}\ and\ \citenamefont
		{Filippov}(2017)}]{Filippov2017}%
	\BibitemOpen
	\bibfield  {author} {\bibinfo {author} {\bibfnamefont {I.~A.}\ \bibnamefont
			{Luchnikov}}\ and\ \bibinfo {author} {\bibfnamefont {S.~N.}\ \bibnamefont
			{Filippov}},\ }\href@noop {} {\bibfield  {journal} {\bibinfo  {journal}
			{Phys. Rev. A}\ }\textbf {\bibinfo {volume} {95}},\ \bibinfo {pages} {022113}
		(\bibinfo {year} {2017})}\BibitemShut {NoStop}%
	\bibitem [{\citenamefont {Wu}\ \emph {et~al.}(2004)\citenamefont {Wu},
		\citenamefont {Lidar},\ and\ \citenamefont
		{Schneider}}]{Wu_entanglement_generation}%
	\BibitemOpen
	\bibfield  {author} {\bibinfo {author} {\bibfnamefont {L.-A.}\ \bibnamefont
			{Wu}}, \bibinfo {author} {\bibfnamefont {D.~A.}\ \bibnamefont {Lidar}}, \
		and\ \bibinfo {author} {\bibfnamefont {S.}~\bibnamefont {Schneider}},\
	}\href@noop {} {\bibfield  {journal} {\bibinfo  {journal} {Phys. Rev. A}\
		}\textbf {\bibinfo {volume} {70}},\ \bibinfo {pages} {032322} (\bibinfo
		{year} {2004})}\BibitemShut {NoStop}%
	\bibitem [{\citenamefont {Pyshkin}\ \emph {et~al.}(2016)\citenamefont
		{Pyshkin}, \citenamefont {Sherman}, \citenamefont {Luo}, \citenamefont
		{You},\ and\ \citenamefont {Wu}}]{Pyshkin_compression}%
	\BibitemOpen
	\bibfield  {author} {\bibinfo {author} {\bibfnamefont {P.~V.}\ \bibnamefont
			{Pyshkin}}, \bibinfo {author} {\bibfnamefont {E.~Y.}\ \bibnamefont
			{Sherman}}, \bibinfo {author} {\bibfnamefont {D.-W.}\ \bibnamefont {Luo}},
		\bibinfo {author} {\bibfnamefont {J.~Q.}\ \bibnamefont {You}}, \ and\
		\bibinfo {author} {\bibfnamefont {L.-A.}\ \bibnamefont {Wu}},\ }\href@noop {}
	{\bibfield  {journal} {\bibinfo  {journal} {Phys. Rev. B}\ }\textbf {\bibinfo
			{volume} {94}},\ \bibinfo {pages} {134313} (\bibinfo {year}
		{2016})}\BibitemShut {NoStop}%
	\bibitem [{\citenamefont {Torres}\ \emph {et~al.}(2017)\citenamefont {Torres},
		\citenamefont {Bern\'ad}, \citenamefont {Alber}, \citenamefont {K\'alm\'an},\
		and\ \citenamefont {Kiss}}]{measurement-chaos-Kiss-2017}%
	\BibitemOpen
	\bibfield  {author} {\bibinfo {author} {\bibfnamefont {J.~M.}\ \bibnamefont
			{Torres}}, \bibinfo {author} {\bibfnamefont {J.~Z.}\ \bibnamefont
			{Bern\'ad}}, \bibinfo {author} {\bibfnamefont {G.}~\bibnamefont {Alber}},
		\bibinfo {author} {\bibfnamefont {O.}~\bibnamefont {K\'alm\'an}}, \ and\
		\bibinfo {author} {\bibfnamefont {T.}~\bibnamefont {Kiss}},\ }\href@noop {}
	{\bibfield  {journal} {\bibinfo  {journal} {Phys. Rev. A}\ }\textbf {\bibinfo
			{volume} {95}},\ \bibinfo {pages} {023828} (\bibinfo {year}
		{2017})}\BibitemShut {NoStop}%
	\bibitem [{\citenamefont {Verstraete}\ \emph {et~al.}(2009)\citenamefont
		{Verstraete}, \citenamefont {Wolf},\ and\ \citenamefont
		{Cirac}}]{Dissipative-control-Verstraete2009}%
	\BibitemOpen
	\bibfield  {author} {\bibinfo {author} {\bibfnamefont {F.}~\bibnamefont
			{Verstraete}}, \bibinfo {author} {\bibfnamefont {M.~M.}\ \bibnamefont
			{Wolf}}, \ and\ \bibinfo {author} {\bibfnamefont {J.~I.}\ \bibnamefont
			{Cirac}},\ }\href@noop {} {\bibfield  {journal} {\bibinfo  {journal} {Nature
				Physics}\ }\textbf {\bibinfo {volume} {5}},\ \bibinfo {pages} {633} (\bibinfo
		{year} {2009})}\BibitemShut {NoStop}%
	\bibitem [{\citenamefont {Jing}\ and\ \citenamefont
		{Yu}(2010)}]{non-markov-Jun-Ting}%
	\BibitemOpen
	\bibfield  {author} {\bibinfo {author} {\bibfnamefont {J.}~\bibnamefont
			{Jing}}\ and\ \bibinfo {author} {\bibfnamefont {T.}~\bibnamefont {Yu}},\
	}\href@noop {} {\bibfield  {journal} {\bibinfo  {journal} {Phys. Rev. Lett.}\
		}\textbf {\bibinfo {volume} {105}},\ \bibinfo {pages} {240403} (\bibinfo
		{year} {2010})}\BibitemShut {NoStop}%
	\bibitem [{\citenamefont {Luo}\ \emph {et~al.}(2015)\citenamefont {Luo},
		\citenamefont {Pyshkin}, \citenamefont {Lam}, \citenamefont {Yu},
		\citenamefont {Lin}, \citenamefont {You},\ and\ \citenamefont
		{Wu}}]{non-markov-Luo}%
	\BibitemOpen
	\bibfield  {author} {\bibinfo {author} {\bibfnamefont {D.-W.}\ \bibnamefont
			{Luo}}, \bibinfo {author} {\bibfnamefont {P.~V.}\ \bibnamefont {Pyshkin}},
		\bibinfo {author} {\bibfnamefont {C.-H.}\ \bibnamefont {Lam}}, \bibinfo
		{author} {\bibfnamefont {T.}~\bibnamefont {Yu}}, \bibinfo {author}
		{\bibfnamefont {H.-Q.}\ \bibnamefont {Lin}}, \bibinfo {author} {\bibfnamefont
			{J.~Q.}\ \bibnamefont {You}}, \ and\ \bibinfo {author} {\bibfnamefont
			{L.-A.}\ \bibnamefont {Wu}},\ }\href@noop {} {\bibfield  {journal} {\bibinfo
			{journal} {Phys. Rev. A}\ }\textbf {\bibinfo {volume} {92}},\ \bibinfo
		{pages} {062127} (\bibinfo {year} {2015})}\BibitemShut {NoStop}%
	\bibitem [{\citenamefont {Schmidt}\ \emph {et~al.}(2011)\citenamefont
		{Schmidt}, \citenamefont {Negretti}, \citenamefont {Ankerhold}, \citenamefont
		{Calarco},\ and\ \citenamefont {Stockburger}}]{Dissipative_sys_control}%
	\BibitemOpen
	\bibfield  {author} {\bibinfo {author} {\bibfnamefont {R.}~\bibnamefont
			{Schmidt}}, \bibinfo {author} {\bibfnamefont {A.}~\bibnamefont {Negretti}},
		\bibinfo {author} {\bibfnamefont {J.}~\bibnamefont {Ankerhold}}, \bibinfo
		{author} {\bibfnamefont {T.}~\bibnamefont {Calarco}}, \ and\ \bibinfo
		{author} {\bibfnamefont {J.~T.}\ \bibnamefont {Stockburger}},\ }\href@noop {}
	{\bibfield  {journal} {\bibinfo  {journal} {Phys. Rev. Lett.}\ }\textbf
		{\bibinfo {volume} {107}},\ \bibinfo {pages} {130404} (\bibinfo {year}
		{2011})}\BibitemShut {NoStop}%
	\bibitem [{\citenamefont {Born}\ and\ \citenamefont {Fock}(1928)}]{Born1928}%
	\BibitemOpen
	\bibfield  {author} {\bibinfo {author} {\bibfnamefont {M.}~\bibnamefont
			{Born}}\ and\ \bibinfo {author} {\bibfnamefont {V.}~\bibnamefont {Fock}},\
	}\href@noop {} {\bibfield  {journal} {\bibinfo  {journal} {Zeitschrift
				f\"{u}r Physik}\ }\textbf {\bibinfo {volume} {51}},\ \bibinfo {pages} {165}
		(\bibinfo {year} {1928})}\BibitemShut {NoStop}%
	\bibitem [{\citenamefont {Torrontegui}\ \emph {et~al.}(2013)\citenamefont
		{Torrontegui}, \citenamefont {Ib{\'{a}}{\~{n}}ez}, \citenamefont
		{Mart{\'{\i}}nez-Garaot}, \citenamefont {Modugno}, \citenamefont {del Campo},
		\citenamefont {Gu{\'{e}}ry-Odelin}, \citenamefont {Ruschhaupt}, \citenamefont
		{Chen},\ and\ \citenamefont {Muga}}]{Shortcuts-Torrontegui-Muga-2013}%
	\BibitemOpen
	\bibfield  {author} {\bibinfo {author} {\bibfnamefont {E.}~\bibnamefont
			{Torrontegui}}, \bibinfo {author} {\bibfnamefont {S.}~\bibnamefont
			{Ib{\'{a}}{\~{n}}ez}}, \bibinfo {author} {\bibfnamefont {S.}~\bibnamefont
			{Mart{\'{\i}}nez-Garaot}}, \bibinfo {author} {\bibfnamefont {M.}~\bibnamefont
			{Modugno}}, \bibinfo {author} {\bibfnamefont {A.}~\bibnamefont {del Campo}},
		\bibinfo {author} {\bibfnamefont {D.}~\bibnamefont {Gu{\'{e}}ry-Odelin}},
		\bibinfo {author} {\bibfnamefont {A.}~\bibnamefont {Ruschhaupt}}, \bibinfo
		{author} {\bibfnamefont {X.}~\bibnamefont {Chen}}, \ and\ \bibinfo {author}
		{\bibfnamefont {J.~G.}\ \bibnamefont {Muga}},\ }in\ \href@noop {} {\emph
		{\bibinfo {booktitle} {Advances In Atomic, Molecular, and Optical Physics}}}\
	(\bibinfo  {publisher} {Elsevier},\ \bibinfo {year} {2013})\ p.\ \bibinfo
	{pages} {117}\BibitemShut {NoStop}%
	\bibitem [{\citenamefont {Gu\'ery-Odelin}\ \emph {et~al.}(2019)\citenamefont
		{Gu\'ery-Odelin}, \citenamefont {Ruschhaupt}, \citenamefont {Kiely},
		\citenamefont {Torrontegui}, \citenamefont {Mart\'{\i}nez-Garaot},\ and\
		\citenamefont {Muga}}]{Muga-review-journal}%
	\BibitemOpen
	\bibfield  {author} {\bibinfo {author} {\bibfnamefont {D.}~\bibnamefont
			{Gu\'ery-Odelin}}, \bibinfo {author} {\bibfnamefont {A.}~\bibnamefont
			{Ruschhaupt}}, \bibinfo {author} {\bibfnamefont {A.}~\bibnamefont {Kiely}},
		\bibinfo {author} {\bibfnamefont {E.}~\bibnamefont {Torrontegui}}, \bibinfo
		{author} {\bibfnamefont {S.}~\bibnamefont {Mart\'{\i}nez-Garaot}}, \ and\
		\bibinfo {author} {\bibfnamefont {J.~G.}\ \bibnamefont {Muga}},\ }\href@noop
	{} {\bibfield  {journal} {\bibinfo  {journal} {Rev. Mod. Phys.}\ }\textbf
		{\bibinfo {volume} {91}},\ \bibinfo {pages} {045001} (\bibinfo {year}
		{2019})}\BibitemShut {NoStop}%
	\bibitem [{\citenamefont {Demirplak}\ and\ \citenamefont
		{Rice}(2003)}]{Demirplak_adiab_drive}%
	\BibitemOpen
	\bibfield  {author} {\bibinfo {author} {\bibfnamefont {M.}~\bibnamefont
			{Demirplak}}\ and\ \bibinfo {author} {\bibfnamefont {S.~A.}\ \bibnamefont
			{Rice}},\ }\href@noop {} {\bibfield  {journal} {\bibinfo  {journal} {The
				Journal of Physical Chemistry A}\ }\textbf {\bibinfo {volume} {107}},\
		\bibinfo {pages} {9937} (\bibinfo {year} {2003})}\BibitemShut {NoStop}%
	\bibitem [{\citenamefont {Berry}(2009)}]{Berry-2009}%
	\BibitemOpen
	\bibfield  {author} {\bibinfo {author} {\bibfnamefont {M.~V.}\ \bibnamefont
			{Berry}},\ }\href@noop {} {\bibfield  {journal} {\bibinfo  {journal} {Journal
				of Physics A: Mathematical and Theoretical}\ }\textbf {\bibinfo {volume}
			{42}},\ \bibinfo {pages} {365303} (\bibinfo {year} {2009})}\BibitemShut
	{NoStop}%
	\bibitem [{\citenamefont {Ren}\ \emph {et~al.}(2017)\citenamefont {Ren},
		\citenamefont {Wang},\ and\ \citenamefont {Gu}}]{Spin-cutting-PLA-Ren2017}%
	\BibitemOpen
	\bibfield  {author} {\bibinfo {author} {\bibfnamefont {F.-H.}\ \bibnamefont
			{Ren}}, \bibinfo {author} {\bibfnamefont {Z.-M.}\ \bibnamefont {Wang}}, \
		and\ \bibinfo {author} {\bibfnamefont {Y.-J.}\ \bibnamefont {Gu}},\
	}\href@noop {} {\bibfield  {journal} {\bibinfo  {journal} {Physics Letters
				A}\ }\textbf {\bibinfo {volume} {381}},\ \bibinfo {pages} {70} (\bibinfo
		{year} {2017})}\BibitemShut {NoStop}%
	\bibitem [{\citenamefont {Loss}\ and\ \citenamefont
		{DiVincenzo}(1998)}]{Loss1998}%
	\BibitemOpen
	\bibfield  {author} {\bibinfo {author} {\bibfnamefont {D.}~\bibnamefont
			{Loss}}\ and\ \bibinfo {author} {\bibfnamefont {D.~P.}\ \bibnamefont
			{DiVincenzo}},\ }\href@noop {} {\bibfield  {journal} {\bibinfo  {journal}
			{Phys. Rev. A}\ }\textbf {\bibinfo {volume} {57}},\ \bibinfo {pages} {120}
		(\bibinfo {year} {1998})}\BibitemShut {NoStop}%
	\bibitem [{\citenamefont {Burkard}\ \emph {et~al.}(1999)\citenamefont
		{Burkard}, \citenamefont {Loss},\ and\ \citenamefont
		{DiVincenzo}}]{Burkard1999}%
	\BibitemOpen
	\bibfield  {author} {\bibinfo {author} {\bibfnamefont {G.}~\bibnamefont
			{Burkard}}, \bibinfo {author} {\bibfnamefont {D.}~\bibnamefont {Loss}}, \
		and\ \bibinfo {author} {\bibfnamefont {D.~P.}\ \bibnamefont {DiVincenzo}},\
	}\href@noop {} {\bibfield  {journal} {\bibinfo  {journal} {Phys. Rev. B}\
		}\textbf {\bibinfo {volume} {59}},\ \bibinfo {pages} {2070} (\bibinfo {year}
		{1999})}\BibitemShut {NoStop}%
	\bibitem [{\citenamefont {Das}\ and\ \citenamefont
		{Chakrabarti}(2008)}]{Das2008}%
	\BibitemOpen
	\bibfield  {author} {\bibinfo {author} {\bibfnamefont {A.}~\bibnamefont
			{Das}}\ and\ \bibinfo {author} {\bibfnamefont {B.~K.}\ \bibnamefont
			{Chakrabarti}},\ }\href@noop {} {\bibfield  {journal} {\bibinfo  {journal}
			{Rev. Mod. Phys.}\ }\textbf {\bibinfo {volume} {80}},\ \bibinfo {pages}
		{1061} (\bibinfo {year} {2008})}\BibitemShut {NoStop}%
	\bibitem [{\citenamefont {McGeoch}(2014)}]{d-wave-book}%
	\BibitemOpen
	\bibfield  {author} {\bibinfo {author} {\bibfnamefont {C.~C.}\ \bibnamefont
			{McGeoch}},\ }\href@noop {} {\emph {\bibinfo {title} {Adiabatic Quantum
				Computation and Quantum Annealing: Theory and Practice (Synthesis Lectures on
				Quantum Computing)}}}\ (\bibinfo  {publisher} {Morgan \& Claypool},\ \bibinfo
	{year} {2014})\BibitemShut {NoStop}%
	\bibitem [{\citenamefont {Schirmer}\ \emph {et~al.}(2002)\citenamefont
		{Schirmer}, \citenamefont {Solomon},\ and\ \citenamefont
		{Leahy}}]{controllability-3-Schirmer2002}%
	\BibitemOpen
	\bibfield  {author} {\bibinfo {author} {\bibfnamefont {S.~G.}\ \bibnamefont
			{Schirmer}}, \bibinfo {author} {\bibfnamefont {A.~I.}\ \bibnamefont
			{Solomon}}, \ and\ \bibinfo {author} {\bibfnamefont {J.~V.}\ \bibnamefont
			{Leahy}},\ }\href@noop {} {\bibfield  {journal} {\bibinfo  {journal} {Journal
				of Physics A: Mathematical and General}\ }\textbf {\bibinfo {volume} {35}},\
		\bibinfo {pages} {4125} (\bibinfo {year} {2002})}\BibitemShut {NoStop}%
	\bibitem [{\citenamefont {Ramakrishna}\ \emph {et~al.}(1995)\citenamefont
		{Ramakrishna}, \citenamefont {Salapaka}, \citenamefont {Dahleh},
		\citenamefont {Rabitz},\ and\ \citenamefont {Peirce}}]{Controllability-2}%
	\BibitemOpen
	\bibfield  {author} {\bibinfo {author} {\bibfnamefont {V.}~\bibnamefont
			{Ramakrishna}}, \bibinfo {author} {\bibfnamefont {M.~V.}\ \bibnamefont
			{Salapaka}}, \bibinfo {author} {\bibfnamefont {M.}~\bibnamefont {Dahleh}},
		\bibinfo {author} {\bibfnamefont {H.}~\bibnamefont {Rabitz}}, \ and\ \bibinfo
		{author} {\bibfnamefont {A.}~\bibnamefont {Peirce}},\ }\href@noop {}
	{\bibfield  {journal} {\bibinfo  {journal} {Phys. Rev. A}\ }\textbf {\bibinfo
			{volume} {51}},\ \bibinfo {pages} {960} (\bibinfo {year} {1995})}\BibitemShut
	{NoStop}%
	\bibitem [{\citenamefont {Burgarth}\ \emph {et~al.}(2009)\citenamefont
		{Burgarth}, \citenamefont {Bose}, \citenamefont {Bruder},\ and\ \citenamefont
		{Giovannetti}}]{local_controllability-1}%
	\BibitemOpen
	\bibfield  {author} {\bibinfo {author} {\bibfnamefont {D.}~\bibnamefont
			{Burgarth}}, \bibinfo {author} {\bibfnamefont {S.}~\bibnamefont {Bose}},
		\bibinfo {author} {\bibfnamefont {C.}~\bibnamefont {Bruder}}, \ and\ \bibinfo
		{author} {\bibfnamefont {V.}~\bibnamefont {Giovannetti}},\ }\href@noop {}
	{\bibfield  {journal} {\bibinfo  {journal} {Phys. Rev. A}\ }\textbf {\bibinfo
			{volume} {79}},\ \bibinfo {pages} {060305(R)} (\bibinfo {year}
		{2009})}\BibitemShut {NoStop}%
	\bibitem [{\citenamefont {Caneva}\ \emph
		{et~al.}(2011{\natexlab{b}})\citenamefont {Caneva}, \citenamefont {Calarco},\
		and\ \citenamefont {Montangero}}]{CRAB-2-Ref16}%
	\BibitemOpen
	\bibfield  {author} {\bibinfo {author} {\bibfnamefont {T.}~\bibnamefont
			{Caneva}}, \bibinfo {author} {\bibfnamefont {T.}~\bibnamefont {Calarco}}, \
		and\ \bibinfo {author} {\bibfnamefont {S.}~\bibnamefont {Montangero}},\
	}\href@noop {} {\bibfield  {journal} {\bibinfo  {journal} {Phys. Rev. A}\
		}\textbf {\bibinfo {volume} {84}},\ \bibinfo {pages} {022326} (\bibinfo
		{year} {2011}{\natexlab{b}})}\BibitemShut {NoStop}%
	\bibitem [{\citenamefont {Rach}\ \emph {et~al.}(2015)\citenamefont {Rach},
		\citenamefont {M\"uller}, \citenamefont {Calarco},\ and\ \citenamefont
		{Montangero}}]{CRAB-3-Ref17}%
	\BibitemOpen
	\bibfield  {author} {\bibinfo {author} {\bibfnamefont {N.}~\bibnamefont
			{Rach}}, \bibinfo {author} {\bibfnamefont {M.~M.}\ \bibnamefont {M\"uller}},
		\bibinfo {author} {\bibfnamefont {T.}~\bibnamefont {Calarco}}, \ and\
		\bibinfo {author} {\bibfnamefont {S.}~\bibnamefont {Montangero}},\
	}\href@noop {} {\bibfield  {journal} {\bibinfo  {journal} {Phys. Rev. A}\
		}\textbf {\bibinfo {volume} {92}},\ \bibinfo {pages} {062343} (\bibinfo
		{year} {2015})}\BibitemShut {NoStop}%
	\bibitem [{\citenamefont {Sakurai}\ and\ \citenamefont
		{Napolitano}(2010)}]{book_sakurai}%
	\BibitemOpen
	\bibfield  {author} {\bibinfo {author} {\bibfnamefont {J.~J.}\ \bibnamefont
			{Sakurai}}\ and\ \bibinfo {author} {\bibfnamefont {J.~J.}\ \bibnamefont
			{Napolitano}},\ }\href@noop {} {\emph {\bibinfo {title} {Modern Quantum
				Mechanics (2nd Edition)}}}\ (\bibinfo  {publisher} {Pearson},\ \bibinfo
	{year} {2010})\BibitemShut {NoStop}%
	\bibitem [{201()}]{2019-1}%
	\BibitemOpen
	\href@noop {} {}\bibinfo {note} {In other words, we change the real function
		$G(\mathbf{a}_{\rm opt},T)=T/2+2a_{\rm 1opt}T/\pi$ (which as we assume must
		have a property $\lim_{T\rightarrow0}G(\mathbf{a}_{\rm opt},T)=0$) to a
		constant~$G(\mathbf{a}_{\rm opt}, T)=0$. Thus the
		factor~$\exp(-iVG(T))=\mathbb{I}$ in Eq.(7)}\BibitemShut {NoStop}%
	\bibitem [{\citenamefont {Pfeuty}(1970)}]{Ising-exact}%
	\BibitemOpen
	\bibfield  {author} {\bibinfo {author} {\bibfnamefont {P.}~\bibnamefont
			{Pfeuty}},\ }\href@noop {} {\bibfield  {journal} {\bibinfo  {journal} {Annals
				of Physics}\ }\textbf {\bibinfo {volume} {57}},\ \bibinfo {pages} {79 }
		(\bibinfo {year} {1970})}\BibitemShut {NoStop}%
	\bibitem [{\citenamefont {Fletcher}(1988)}]{BFGS}%
	\BibitemOpen
	\bibfield  {author} {\bibinfo {author} {\bibfnamefont {R.}~\bibnamefont
			{Fletcher}},\ }\href@noop {} {\emph {\bibinfo {title} {Practical Methods of
				Optimization, 2nd Edition}}}\ (\bibinfo  {publisher} {Wiley},\ \bibinfo
	{year} {1988})\BibitemShut {NoStop}%
	\bibitem [{\citenamefont {Mandelstam}\ and\ \citenamefont
		{Tamm}(1991)}]{Mandelstam1991}%
	\BibitemOpen
	\bibfield  {author} {\bibinfo {author} {\bibfnamefont {L.}~\bibnamefont
			{Mandelstam}}\ and\ \bibinfo {author} {\bibfnamefont {I.}~\bibnamefont
			{Tamm}},\ }in\ \href@noop {} {\emph {\bibinfo {booktitle} {Selected
				Papers}}},\ \bibinfo {editor} {edited by\ \bibinfo {editor} {\bibfnamefont
			{B.~M.}\ \bibnamefont {Bolotovskii}}, \bibinfo {editor} {\bibfnamefont
			{V.~Y.}\ \bibnamefont {Frenkel}}, \ and\ \bibinfo {editor} {\bibfnamefont
			{R.}~\bibnamefont {Peierls}}}\ (\bibinfo  {publisher} {Springer Berlin
		Heidelberg},\ \bibinfo {address} {Berlin, Heidelberg},\ \bibinfo {year}
	{1991})\ p.\ \bibinfo {pages} {115}\BibitemShut {NoStop}%
	\bibitem [{\citenamefont {Margolus}\ and\ \citenamefont
		{Levitin}(1998)}]{MARGOLUS1998188}%
	\BibitemOpen
	\bibfield  {author} {\bibinfo {author} {\bibfnamefont {N.}~\bibnamefont
			{Margolus}}\ and\ \bibinfo {author} {\bibfnamefont {L.~B.}\ \bibnamefont
			{Levitin}},\ }\href@noop {} {\bibfield  {journal} {\bibinfo  {journal}
			{Physica D: Nonlinear Phenomena}\ }\textbf {\bibinfo {volume} {120}},\
		\bibinfo {pages} {188} (\bibinfo {year} {1998})}\BibitemShut {NoStop}%
	\bibitem [{\citenamefont {Deffner}\ and\ \citenamefont
		{Lutz}(2013)}]{QSL_Deffner_2013}%
	\BibitemOpen
	\bibfield  {author} {\bibinfo {author} {\bibfnamefont {S.}~\bibnamefont
			{Deffner}}\ and\ \bibinfo {author} {\bibfnamefont {E.}~\bibnamefont {Lutz}},\
	}\href@noop {} {\bibfield  {journal} {\bibinfo  {journal} {Journal of Physics
				A: Mathematical and Theoretical}\ }\textbf {\bibinfo {volume} {46}},\
		\bibinfo {pages} {335302} (\bibinfo {year} {2013})}\BibitemShut {NoStop}%
	\bibitem [{\citenamefont {Murphy}\ \emph {et~al.}(2010)\citenamefont {Murphy},
		\citenamefont {Montangero}, \citenamefont {Giovannetti},\ and\ \citenamefont
		{Calarco}}]{QSL-in-chain-Murphy-Ref15}%
	\BibitemOpen
	\bibfield  {author} {\bibinfo {author} {\bibfnamefont {M.}~\bibnamefont
			{Murphy}}, \bibinfo {author} {\bibfnamefont {S.}~\bibnamefont {Montangero}},
		\bibinfo {author} {\bibfnamefont {V.}~\bibnamefont {Giovannetti}}, \ and\
		\bibinfo {author} {\bibfnamefont {T.}~\bibnamefont {Calarco}},\ }\href@noop
	{} {\bibfield  {journal} {\bibinfo  {journal} {Phys. Rev. A}\ }\textbf
		{\bibinfo {volume} {82}},\ \bibinfo {pages} {022318} (\bibinfo {year}
		{2010})}\BibitemShut {NoStop}%
	\bibitem [{\citenamefont {Caneva}\ \emph {et~al.}(2009)\citenamefont {Caneva},
		\citenamefont {Murphy}, \citenamefont {Calarco}, \citenamefont {Fazio},
		\citenamefont {Montangero}, \citenamefont {Giovannetti},\ and\ \citenamefont
		{Santoro}}]{QSL-quant-contr}%
	\BibitemOpen
	\bibfield  {author} {\bibinfo {author} {\bibfnamefont {T.}~\bibnamefont
			{Caneva}}, \bibinfo {author} {\bibfnamefont {M.}~\bibnamefont {Murphy}},
		\bibinfo {author} {\bibfnamefont {T.}~\bibnamefont {Calarco}}, \bibinfo
		{author} {\bibfnamefont {R.}~\bibnamefont {Fazio}}, \bibinfo {author}
		{\bibfnamefont {S.}~\bibnamefont {Montangero}}, \bibinfo {author}
		{\bibfnamefont {V.}~\bibnamefont {Giovannetti}}, \ and\ \bibinfo {author}
		{\bibfnamefont {G.~E.}\ \bibnamefont {Santoro}},\ }\href@noop {} {\bibfield
		{journal} {\bibinfo  {journal} {Phys. Rev. Lett.}\ }\textbf {\bibinfo
			{volume} {103}},\ \bibinfo {pages} {240501} (\bibinfo {year}
		{2009})}\BibitemShut {NoStop}%
	\bibitem [{\citenamefont {Pyshkin}\ \emph {et~al.}(2018)\citenamefont
		{Pyshkin}, \citenamefont {Sherman}, \citenamefont {You},\ and\ \citenamefont
		{Wu}}]{Pyshkin2018-NJP}%
	\BibitemOpen
	\bibfield  {author} {\bibinfo {author} {\bibfnamefont {P.~V.}\ \bibnamefont
			{Pyshkin}}, \bibinfo {author} {\bibfnamefont {E.~Y.}\ \bibnamefont
			{Sherman}}, \bibinfo {author} {\bibfnamefont {J.~Q.}\ \bibnamefont {You}}, \
		and\ \bibinfo {author} {\bibfnamefont {L.-A.}\ \bibnamefont {Wu}},\
	}\href@noop {} {\bibfield  {journal} {\bibinfo  {journal} {New Journal of
				Physics}\ }\textbf {\bibinfo {volume} {20}},\ \bibinfo {pages} {105006}
		(\bibinfo {year} {2018})}\BibitemShut {NoStop}%
	\bibitem [{\citenamefont {Pyshkin}\ \emph {et~al.}(2019)\citenamefont
		{Pyshkin}, \citenamefont {Sherman},\ and\ \citenamefont {Wu}}]{Acta2019}%
	\BibitemOpen
	\bibfield  {author} {\bibinfo {author} {\bibfnamefont {P.}~\bibnamefont
			{Pyshkin}}, \bibinfo {author} {\bibfnamefont {E.}~\bibnamefont {Sherman}}, \
		and\ \bibinfo {author} {\bibfnamefont {L.-A.}\ \bibnamefont {Wu}},\
	}\href@noop {} {\bibfield  {journal} {\bibinfo  {journal} {Acta Physica
				Polonica A}\ }\textbf {\bibinfo {volume} {135}},\ \bibinfo {pages} {1198}
		(\bibinfo {year} {2019})}\BibitemShut {NoStop}%
	\bibitem [{\citenamefont {Rabitz}\ \emph {et~al.}(2004)\citenamefont {Rabitz},
		\citenamefont {Hsieh},\ and\ \citenamefont
		{Rosenthal}}]{easy-control-Rabitz1998}%
	\BibitemOpen
	\bibfield  {author} {\bibinfo {author} {\bibfnamefont {H.~A.}\ \bibnamefont
			{Rabitz}}, \bibinfo {author} {\bibfnamefont {M.~M.}\ \bibnamefont {Hsieh}}, \
		and\ \bibinfo {author} {\bibfnamefont {C.~M.}\ \bibnamefont {Rosenthal}},\
	}\href@noop {} {\bibfield  {journal} {\bibinfo  {journal} {Science}\ }\textbf
		{\bibinfo {volume} {303}},\ \bibinfo {pages} {1998} (\bibinfo {year}
		{2004})}\BibitemShut {NoStop}%
	\bibitem [{\citenamefont {Ho}\ and\ \citenamefont
		{Rabitz}(2006)}]{easy-to-find-control}%
	\BibitemOpen
	\bibfield  {author} {\bibinfo {author} {\bibfnamefont {T.-S.}\ \bibnamefont
			{Ho}}\ and\ \bibinfo {author} {\bibfnamefont {H.}~\bibnamefont {Rabitz}},\
	}\href@noop {} {\bibfield  {journal} {\bibinfo  {journal} {Journal of
				Photochemistry and Photobiology A: Chemistry}\ }\textbf {\bibinfo {volume}
			{180}},\ \bibinfo {pages} {226 } (\bibinfo {year} {2006})}\BibitemShut
	{NoStop}%
	\bibitem [{\citenamefont {Pechen}\ and\ \citenamefont
		{Tannor}(2011)}]{are-there-traps-in-landscapes}%
	\BibitemOpen
	\bibfield  {author} {\bibinfo {author} {\bibfnamefont {A.~N.}\ \bibnamefont
			{Pechen}}\ and\ \bibinfo {author} {\bibfnamefont {D.~J.}\ \bibnamefont
			{Tannor}},\ }\href@noop {} {\bibfield  {journal} {\bibinfo  {journal} {Phys.
				Rev. Lett.}\ }\textbf {\bibinfo {volume} {106}},\ \bibinfo {pages} {120402}
		(\bibinfo {year} {2011})}\BibitemShut {NoStop}%
	\bibitem [{\citenamefont {Zhdanov}\ and\ \citenamefont
		{Seideman}(2015)}]{role-of-constraints-in-control}%
	\BibitemOpen
	\bibfield  {author} {\bibinfo {author} {\bibfnamefont {D.~V.}\ \bibnamefont
			{Zhdanov}}\ and\ \bibinfo {author} {\bibfnamefont {T.}~\bibnamefont
			{Seideman}},\ }\href@noop {} {\bibfield  {journal} {\bibinfo  {journal}
			{Phys. Rev. A}\ }\textbf {\bibinfo {volume} {92}},\ \bibinfo {pages} {052109}
		(\bibinfo {year} {2015})}\BibitemShut {NoStop}%
	\bibitem [{\citenamefont {Zhdanov}(2018)}]{easy-control-criticizm-Zhdanov2018}%
	\BibitemOpen
	\bibfield  {author} {\bibinfo {author} {\bibfnamefont {D.~V.}\ \bibnamefont
			{Zhdanov}},\ }\href@noop {} {\bibfield  {journal} {\bibinfo  {journal}
			{Journal of Physics A: Mathematical and Theoretical}\ }\textbf {\bibinfo
			{volume} {51}},\ \bibinfo {pages} {508001} (\bibinfo {year}
		{2018})}\BibitemShut {NoStop}%
\end{thebibliography}


\end{document}